%% file: main.tex
\documentclass[sigconf]{acmart}
\input{preamble/packages}
\input{preamble/definitions}

\copyrightyear{2025}
\acmYear{2025}
\setcopyright{cc}
\setcctype{by}
\acmConference[ICTIR '25]{Proceedings of the 2025 International ACM SIGIR
Conference on Innovative Concepts and Theories in Information Retrieval
(ICTIR)}{July 18, 2025}{Padua, Italy}
\acmBooktitle{Proceedings of the 2025 International ACM SIGIR Conference on
Innovative Concepts and Theories in Information Retrieval (ICTIR) (ICTIR '25), July
18, 2025, Padua, Italy}\acmDOI{10.1145/3731120.3744620}
\acmISBN{979-8-4007-1861-8/2025/07}

\begin{document}

\title[Towards Fair Rankings: Leveraging LLMs for Gender Bias Detection and Measurement]{Towards Fair Rankings: Leveraging LLMs\\ for Gender Bias Detection and Measurement}

\author{Maryam Mousavian}
\orcid{https://orcid.org/0009-0001-6283-0355}
\affiliation{%
  \institution{Università della Svizzera italiana  \\ \&  University of Amsterdam}
  \country{}
  }
\email{maryamalsadat.mousavian@usi.ch}

\author{Zahra Abbasiantaeb}
\orcid{https://orcid.org/0000-0002-4046-3419}
\affiliation{%
  \institution{University of Amsterdam}
  \country{The Netherland}
  }
\email{z.abbasiantaeb@uva.nl}

\author{Mohammad Aliannejadi}
\orcid{https://orcid.org/0000-0002-9447-4172}
\affiliation{%
  \institution{Unviersity of Amsterdam}
  \country{The Netherland}
  }
\email{m.aliannejadi@uva.nl}

\author{Fabio Crestani}
\orcid{https://orcid.org/0000-0001-8672-0700}
\affiliation{%
  \institution{Università della Svizzera italiana}
  \country{Switzerland}
  }
\email{fabio.crestani@usi.ch}

\renewcommand{\shortauthors}{Mousavian et al.}

\begin{abstract}
The presence of social biases in \ac{NLP} and \ac{IR} systems is an ongoing challenge, which underlines the importance of developing robust approaches to identifying and evaluating such biases. In this paper, we aim to address this issue by leveraging \acp{LLM} to detect and measure gender bias in passage ranking. 
Existing gender fairness metrics rely on lexical- and frequency-based measures, leading to various limitations, e.g., missing subtle gender disparities. Building on our LLM-based gender bias detection method, we introduce a novel gender fairness metric, named \ac{CWEx}, aiming to address existing limitations.
To measure the effectiveness of our proposed metric and study LLMs' effectiveness on detecting gender bias, we annotate a subset of the MS MARCO Passage Ranking collection and release our new gender bias collection, called MSMGenderBias, to foster future research in this area.\footnote{The data can be found here: \url{https://github.com/MaryamMousavian/MSMGenderBias}.} 
Our extensive experimental results on various ranking models show that our proposed metric offers a more detailed evaluation of fairness compared to previous metrics, with improved alignment to human labels (58.77\% for Grep-BiasIR, and 18.51\% for MSMGenderBias, measured using Cohen's $\kappa$ agreement), effectively distinguishing gender bias in ranking. By integrating LLM-driven bias detection, an improved fairness metric, and gender bias annotations for an established dataset, this work provides a more robust framework for analyzing and mitigating bias in \ac{IR} systems.
\end{abstract}

\ccsdesc[500]{Information systems~Retrieval models and ranking}
\ccsdesc[500]{Information systems~Evaluation of retrieval results}

\keywords{Bias, Fairness, Evaluation, Ranking, Large Language Models}

\maketitle

\input{sections/1-intro}
\input{sections/2-related}
\input{sections/3-method}

\input{sections/4-exp-setup}

\input{sections/5-exp-design-result}

\input{sections/6-conclusion}
\input{sections/7-limitations}
\input{sections/8-acknowledgment}

\bibliography{main}
\bibliographystyle{ACM-Reference-Format}
\balance

\appendix
\input{sections/9-appendix}

\end{document}

%% file: preamble/packages.tex
\usepackage{multirow}
\usepackage[]{xcolor}
\usepackage{tabularx}
\usepackage{xspace}
\usepackage{acronym}
\usepackage{arydshln}
\usepackage{subcaption} 

\usepackage{amssymb}
\usepackage{amsmath} 
\usepackage{graphicx}
\usepackage{flushend}
\usepackage[htt]{hyphenat}
\usepackage[inline]{enumitem}
\usepackage{stfloats}

%% file: preamble/definitions.tex
\definecolor{midnightgreen}{rgb}{0.0, 0.29, 0.33}

\newcommand{\nfairr}{NFaiRR\xspace}
\newcommand{\texfair}{TExFAIR\xspace}

\acrodef{AI}{Artificial Intelligence}
\acrodef{LLM}{Large Language Model}
\acrodef{PLM}{Pre-trained Language Model}
\acrodef{IR}{Information Retrieval}
\acrodef{NLP}{Natural Language Processing}
\acrodef{CWEx}{Class-wise Weighted Exposure}
\acrodef{CoT}{Chain of Thought}
\acrodef{NLI}{Natural Language Inference}
\acrodef{WEAT}{Word Embedding Association Test}
\acrodef{SEAT}{Sentence Encoder Association Test}
\acrodef{CEAT}{Contextualized Embedding Association Test}
\acrodef{DisCo}{Discovery of Correlations}
\acrodef{LPBS}{Log-Probability Bias Score}
\acrodef{SGS}{Social Group Substitutions}
\acrodef{DR}{Demographic Representation}
\acrodef{ST}{Stereotypical Associations}
\acrodef{CAT}{Context Association Test}
\acrodef{iCAT}{Idealized CAT}
\acrodef{AUL}{All Unmasked Likelihood}
\acrodef{AULA}{AUL with Attention Weights}
\acrodef{MLM}{Masked Language Model}
\acrodef{LMB}{Language Model Bias}

\newcommand{\header}[1]{\vspace{1mm}\noindent\textbf{#1}}

%% file: sections/1-intro.tex
\section{Introduction}
\label{sec:intro}

\input{fig/bias-exp}

With recent advances of \acfp{LLM} and their widespread adoption, addressing bias and fairness inherent in these models and various \acf{IR} applications has become essential. These systems are often vulnerable to bias and are being used in important areas like healthcare, education, hiring, and legal, where even small biases can lead to serious consequences.
Consequently, both \ac{NLP} and \ac{IR} communities have made significant efforts to mitigate bias and unfairness in their methods, to advance development of more equitable and transparent systems~\cite{mehrabi2021survey, gallegos2024bias}. Effective bias mitigation first requires a thorough detection and evaluation of bias originating from different sources. 

Current gender bias metrics primarily depend on lexical and term-based approaches~\cite{rekabsaz_neural_2020, rekabsaz_societal_2021, abolghasemi_measuring_2024}, leading to an oversimplified definition of gender bias and resulting in various limitations. 
These approaches look for explicit and straightforward references to genders and rely on a limited set of predefined gender-related terms for detecting bias. Hence, they miss more complex and obfuscated biases in text, leaving a big gap between the ability of the existing metrics to detect such biases and the more recent and advanced text generation models. This could lead to various limitations and risks when it comes to gender bias detection, as the potentially biased models are much more capable than the metrics used to detect bias.

Looking at the two example sentences in Figure~\ref{fig:bias-example}, sentence (a) shows a bias toward the male gender because it only includes male-related terms. The existing approaches can easily detect the bias of this sentence because it only contains male-related terms.
In contrast, however, sentence (b) is also biased toward the male gender, but it often goes undetected by existing term-based methods because it contains both male and female terms in phrases like "his mother" and "her son."
With existing bias detection methods, these phrases are often broken into separate words that carry different gender identities, which can misrepresent the actual target of the bias.

To address these limitations, we study the evaluation and measurement of gender bias, with a particular focus on the evaluation of bias in ranked lists.
Ranking models tend to rank the documents in a biased way~\cite{10.1145/3292500.3330691, 10.1145/3173574.3174225, fabris2020gender}, which can unfairly affect disadvantaged social groups based on characteristics like gender, race, ethnicity, and age. 
In particular, gender bias can cause ranking models to systematically favor one gender over another. 
To address the mentioned limitations, we propose leveraging \acp{LLM} to classify documents based on gender for evaluation purposes and investigate the following research questions (RQs):
\begin{enumerate}[label=\textbf{RQ\arabic*}]
    \item : How effective are \acp{LLM} in detecting gender bias of documents? \label{rq1}

    \item : How do different \acp{LLM} perform in gender classification of documents with typical gender stereotypes? \label{rq2}

    \item : How can the gender bias detection capabilities of \acp{LLM} be used to evaluate gender bias in ranked lists? \label{rq3}

\end{enumerate}

To address these research questions, we conducted a series of experiments. 
For \ref{rq1}, we utilize open-source and closed-source \acp{LLM}, to classify documents based on gender bias into three categories, including neutral (no bias), male (biased towards male), and female (biased towards female). To measure the effectiveness of \acp{LLM} in detecting gender bias of documents, we conduct a human study and compare the prediction of \acp{LLM} with human labels.
For \ref{rq2}, we assess the ability of \acp{LLM} to detect gender bias in documents that contain stereotypes related to both male and female genders. Given the presence of gender-related stereotypes, \acp{LLM} may exhibit biased behavior in identifying the true target of the bias. To explore this, we analyze the models' performance in detecting gender bias across both male- and female-related stereotypes, checking whether their effectiveness varies depending on the gender.
Finally, to answer \ref{rq3}, we propose a new metric to measure the fairness of the ranking list of documents, based on gender categories. 
This research approach can be applied not only to gender bias but also to biases involving other social attributes, including race, age, and ethnicity. 

We summarize our contributions in this work as follows.
\begin{itemize}[nosep,leftmargin=*]
    \item We propose a method that uses \ac{LLM} and prompt engineering to detect gender bias in documents, effectively leveraging the linguistic capabilities of \acp{LLM} to reveal subtle and complex forms of bias.
    \item We introduce \ac{CWEx}, a fairness metric designed to quantify gender bias in ranking systems by measuring disparities in gender exposure within ranked lists.
    \item We release MSMGenderBias, a publicly available dataset containing gender bias annotations for a subset of the MS MARCO Passage Ranking collection~\cite{bajaj2016ms}.
    \item We validate the effectiveness of our proposed method for detecting gender bias using \acp{LLM} by evaluating it across multiple ranking models and datasets, including Grep-BiasIR~\cite{krieg2023grep} and MSMGenderBias.
\end{itemize}

%% file: fig/bias-exp.tex
\begin{figure}[t]
  \centering
  \begin{subfigure}[t]{\columnwidth}
      \centering
      \includegraphics[width=\columnwidth, trim=10 27 10 10, clip]{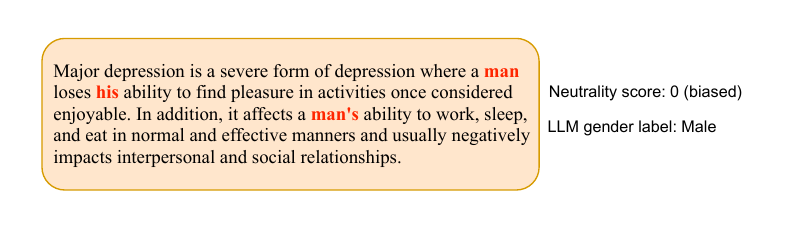}
      \caption{Example passage correctly identified as biased (towards male) by both the lexical-based and LLM-based measures.}
      \label{fig:bias-example-correct}
  \end{subfigure}

  \begin{subfigure}[t]{\columnwidth}
      \centering
      \includegraphics[width=\columnwidth, trim=10 25 10 0, clip]{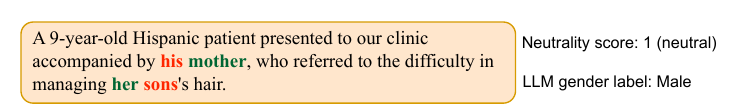}
      \caption{Example passage correctly identified as biased (towards male) by the LLM-based measures, but identified as neutral by the lexical-based measure.}
      \label{fig:bias-example-incorrect}
  \end{subfigure}
  \caption{Example passages labeled for their gender bias, both by the existing lexical-based neutrality metric (used by NFaiRR, and the LLM-based gender bias detection (ours). We see in (a) that, given the simplicity of the language and the explicit use of the gender-related terms, both metrics manage to identify the bias correctly. However, in (b), given that the text is slightly more complex, having gender-related terms from both genders, the lexical-based metric fails to identify the bias in the text and labels it as neutral, while the LLM-based approach is able to assign the correct label.}
  \label{fig:bias-example}
\end{figure}

%% file: sections/2-related.tex
\section{Related Work}\label{sec:related-work}

\header{Gender Bias Evaluation.}
Bias metrics and datasets are inherently tailored to specific tasks, as gender bias can manifest differently across various \ac{IR} and \ac{NLP} tasks. These variations shape how fairness is defined and measured within each context.
In \ac{NLP}, gender bias has been widely studied, with research focusing on its presence in embeddings~\cite{may-etal-2019-measuring, dev-etal-2021-oscar, 10.1145/3461702.3462536}, \ac{LLM}~\cite{liang2021towards}, and downstream applications such as machine translation~\cite{mechura-2022-taxonomy}, question answering ~\cite{li-etal-2020-unqovering, parrish-etal-2022-bbq}, text generation~\cite{sheng-etal-2019-woman}, and \ac{NLI}~\cite{dev2020measuring}. 

While gender bias in \ac{NLP} has been studied across various tasks, increasing focus has recently shifted to its presence in large-scale models such as \acp{LLM}. 
These models serve not only as foundational components in diverse model architectures but also as standalone systems for task execution through prompt engineering. 
Metrics for evaluating gender bias are categorized into three types including embedding-based, probability-based, and generated text-based metrics. 

The embedding-based metrics are classified into word embedding (\ac{WEAT}~\cite{caliskan2017semantics} and \ac{WEAT}*~\cite{dev-etal-2021-oscar}) and sentence embedding metrics (\ac{SEAT}~\cite{may-etal-2019-measuring} and \ac{CEAT}~\cite{10.1145/3461702.3462536}), typically measuring the distance between neutral words and gender-related words in the vector space. 
For static word embeddings, \textbf{\ac{WEAT}} assesses stereotypical associations between social group concepts and neutral attributes using a test statistic modeled on the Implicit Association Test~\cite{greenwald1998measuring}. 
The \textbf{\ac{WEAT}*} metric, a variant of the \ac{WEAT} metric, modifies this approach to identify meaningful associations between male and female terms rather than stereotypical ones. 
For sentence embedding metrics, \ac{SEAT} and \ac{CEAT} were proposed by extending the \ac{WEAT} approach to contextualized embeddings. 
The \textbf{\ac{SEAT}} metric applies the \ac{WEAT} methodology to sentence representations rather than word representations, which means it compares sets of sentences instead of sets of words. 
The \textbf{\ac{CEAT}} metric builds upon the \ac{WEAT} approach by generating sentences through combinations of two sets of target words and two sets of attribute words. 
Then it randomly selects a subset of embeddings and computes a distribution of effect sizes. 
\textbf{Sentence Bias Score}~\cite{dolci2023improving} measures bias by computing the cosine similarity between neutral words and the gender direction, then normalizing the total word-level bias based on sentence length and each word's contextualized semantic significance.

Probability-based metrics, based on model-assigned token probabilities, assess bias using fill-in-the-blank templates or sentence likelihoods with masked token and pseudo-log-likelihood methods. 
The \ac{DisCo}~\cite{webster2020measuring}, \ac{LPBS}~\cite{kurita-etal-2019-measuring}, and Categorical Bias Score~\cite{ahn-oh-2021-mitigating} metrics were introduced as masked token metrics. 
The \textbf{\ac{DisCo}} metric assesses template-based sentence completions by inserting a bias trigger word into one slot and using the model’s top three predictions for the other. 
The score is derived by counting gender-associated completions and averaging across templates. 
The \textbf{\ac{LPBS}} metric, similar to \ac{DisCo}, uses a template-based approach to measure bias in neutral attribute words. 
The metric normalizes a token's predicted probability in "[MASK] is a [NEUTRAL ATTRIBUTE]" using the model's prior probability from "[MASK] is a [MASK]" to remove inherent biases. 
The bias score is then computed as the difference in normalized probabilities between two opposing social groups. 
The \textbf{Categorical Bias Score} is based on the \ac{LPBS} metric and quantifies bias by measuring the variance in predicted tokens for fill-in-the-blank templates across different social groups based on protected attribute words. 
The pseudo-log-likelihood metrics (CrowS-Pairs Score~\cite{nangia-etal-2020-crows}, \ac{CAT}~\cite{nadeem-etal-2021-stereoset}, and \ac{AUL}~\cite{kaneko2022unmasking}) estimate the probability of a token by masking and predicting it based on the remaining unmasked tokens in the sentence~\cite{salazar-etal-2020-masked, wang-cho-2019-bert}. 
The \textbf{CrowS-Pairs Score} measures the percentage of cases where the model assigns a higher (pseudo) likelihood to a stereotypical sentence over a less stereotypical one, estimating $P (U |M, \theta)$ by masking and predicting unmodified tokens. 
In a pair of stereotyped and anti-stereotyped sentences, $U$ represents shared tokens, $M$ consists of modified tokens, and $\theta$ denotes the parameters of the model. 
In contrast, \textbf{\ac{CAT}} estimates $P (M |U, \theta)$ instead of $P (U |M, \theta)$. \textbf{\ac{AUL}} extends the CrowS-Pairs Score by predicting all tokens in the complete sentence, improving prediction accuracy through full context and eliminating selection bias in word masking.

Generated text-based metrics are classified into distribution-, classifier-, and lexicon-based approaches. 
Distribution-based metrics, such as Co-occurrence Bias Score~\cite{bordia-bowman-2019-identifying}, \ac{DR}~\cite{bommasani2023holistic}, and \ac{ST}~\cite{bommasani2023holistic}, detect bias by comparing token distributions between social groups. 
The \textbf{Co-occurrence Bias Score} measures the likelihood of a word appearing with gendered terms, while \textbf{\ac{DR}} metric generates a vector of social group frequencies in LLM-generated text, comparing it to a reference distribution using metrics like KL divergence. 
The \textbf{\ac{ST}} metric, similar to \ac{DR}, measures gender association bias based on associations with specific terms like occupations. 
Classifier-based metrics use auxiliary models to assess toxicity, sentiment, or bias towards social group attributes~\cite{sheng-etal-2019-woman, huang-etal-2020-reducing, sicilia-alikhani-2023-learning}. 
Lexicon-based metrics compare output words against a list of harmful words or assign pre-calculated bias scores~\cite{nozza-etal-2021-honest, 10.1145/3442188.3445924}.

In \ac{IR}, research on gender bias in document rankings has predominantly focused on gender as a binary attribute, examining how male and female groups are represented~\cite{rekabsaz_societal_2021, rekabsaz_neural_2020, zerveas_mitigating_2022, heuss_predictive_2023, bigdeli_-biasing_2023}. 
Existing research on gender bias evaluation~\cite{rekabsaz_neural_2020, rekabsaz_societal_2021, abolghasemi_measuring_2024} has mainly relied on a predefined set of gender-representative words to compute gender bias scores. 
The Average Rank Bias (ARB) metric~\cite{rekabsaz_neural_2020} is designed to provide a comprehensive assessment of gender bias in a model’s document ranking list. 
To incorporate the influence of ranking positions throughout the list, \textbf{ARB} calculates the average gender magnitude of documents at each ranking position for each query.
To address the limitation of the ARB metric, which does not account for background bias in the collection, the \nfairr metric~\cite{rekabsaz_societal_2021} is introduced.
\textbf{\nfairr}, a most recently used fairness measure, evaluates the fairness of retrieved documents by considering both the positions of the document (position bias) and the neutrality of the document (unbiasedness). 
This metric identifies documents that offer a balanced representation concerning a protected attribute. 
The recently proposed fairness metric, \textbf{\texfair}~\cite{abolghasemi_measuring_2024}, extends the AWRF~\cite{sapiezynski_quantifying_2019} evaluation framework by introducing a metric that explicitly defines the association of each document with groups using a probabilistic term-level approach. 
The \texfair is developed to address limitations of the \nfairr metric, where individual documents can exhibit bias even if the overall ranked list maintains a balanced group representation.
In the TREC Fair Ranking Track~\cite{ekstrand2023overview}, AWRF is also applied to the single-ranking task with a specified configuration. Unlike the original AWRF, which supports flexible attention models and distance functions, TREC fixed the attention to a logarithmic decay and used Jensen-Shannon Divergence (JSD). Target group distributions were defined by averaging empirical group distributions among relevant documents with either the world population (for location) or a uniform distribution (for gender). The track also emphasized intersectional fairness by computing exposure over the Cartesian product of multiple group attributes (location and gender) and handled unknown attributes by treating them as separate groups or ignoring documents with fully unknown gender or location. The fairness evaluation metric used in the NTCIR FairWeb Task \cite{tao2025overview} assesses how closely the distribution of groups in the top-ranked results matches a predefined target distribution for each attribute. For nominal attributes (e.g., gender, origin), it employs JSD, while ordinal attributes (e.g., ratings, h-index) are evaluated using metrics such as Normalized Mean Difference (NMD) or Relative Normalized Ordering Difference (RNOD). The observed group distribution at each rank is computed by averaging the group membership vectors of the documents retrieved up to that point.

\header{\acp{LLM} as Gender Bias Evaluator.}
Recent research in \ac{IR} and \ac{NLP} has increasingly utilized \acp{LLM} as evaluators, taking advantage of their alignment with human preferences and beliefs, which influence objective assessments across tasks such as relevance judgment and response generation~\cite{li2024generative, kumar_decoding_2024, abbasiantaeb2024can, 10.1145/3578337.3605136, rahmani2024synthetic}.

To develop effective bias mitigation methods, it is essential to first establish reliable bias evaluation approaches to assess their performance.
Evaluation metrics can focus on intrinsic properties of \acp{LLM} or their extrinsic biased behavior in downstream tasks. 
For instance, a pipeline proposed by~\cite{kumar_decoding_2024} consists of three \acp{LLM}—an attacker, a target, and an evaluator—to detect and assess gender bias in response generation. 
The attacker uses adversarial prompts to induce biased responses from the target \ac{LLM}, aiming to exploit its potential vulnerabilities. 
The generated biased responses are subsequently evaluated by the evaluator, which rates the degree of bias on a scale from 0 (No Bias) to 5 (Extreme Bias). 
The final scores are calculated by averaging the differences in bias scores between the genders for each response.

%% file: sections/3-method.tex
\section{Method}

The existing fairness metrics for ranking rely on the lexical-based methods, which use the count of gender-representative words for gender bias detection. 
As outlined in section \ref{sec:intro}, different from the existing fairness metrics for ranking, our goal is to propose a semantic-based approach to detect and identify gender bias in documents.
To do so, we first leverage the capabilities of \acp{LLM} to detect gender bias and classify documents into three categories: neutral, male, and female, compensating for the lack of large labeled datasets for gender bias detection in ranking. Using these categories, we introduce a new fairness metric. To the best of our knowledge, no existing metric in the literature computes fairness in ranking by leveraging the distribution of documents between the three gender bias categories: male-biased, female-biased, and neutral. The most relevant work, \nfairr (see Section~\ref{sec:related-work}), assigns a neutrality score to each document. However, following their discussion, we interpret these scores to categorize documents as either neutral or non-neutral, enabling us to use \nfairr as a baseline. In contrast, our proposed metric explicitly incorporates the exposure of all three bias categories, allowing for a more fine-grained and interpretable assessment of fairness in ranked outputs. We will explain our bias detection method and fairness metric in the following. 

\subsection{LLMs as Bias Detector}
\label{sec:biasdetect}
To classify documents based on gender bias, we employ both closed-source and open-source \acp{LLM}. 
To identify the bias of documents, we rely on few-shot prompting of the \acp{LLM}.
We employ four different zero-, one-, three-shot, and \acp{CoT} prompts for annotating the bias of documents. 
Our prompts are designed with inspiration from existing related works~\cite{abbasiantaeb2024can, kumar_decoding_2024}. 
The designed prompts are shown in Table \ref{tab:prompt}. A detailed explanation of the designed prompts is provided below.
\begin{itemize}[leftmargin=*]
    \item \textbf{Zero-shot}: The zero-shot prompt serves as the primary prompt in all four settings.
    It includes the detailed instructions given to the \acp{LLM} to classify documents based on gender bias.
    
    \item \textbf{One-shot}: In the one-shot prompt, we augment our zero-shot prompt by one demonstration. The demonstration is an example of a document with a neutral label of gender bias.
    
    \item \textbf{Three-shot}: The three-shot prompt includes three different demonstrations, each corresponding to a different gender bias label, alongside the main prompt.
    
    \item \textbf{\acp{CoT}}: In our \acp{CoT} prompt, we instruct the model to provide reasoning along with the document label.
\end{itemize}

\input{tables/prompt}

\subsection{Group Fairness Metric}
To define a new metric for measuring the fairness of ranked lists using gender bias labels, we take the following essential properties into account.

\begin{enumerate}[leftmargin=*]
    \item Introduce a weight factor to prioritize neutral documents in higher ranks while allowing customization between neutral exposure and gender group disparity, making the metric adaptable to different fairness goals.
    \item Enhance the exposure of neutral documents while reducing exposure disparities between gender groups.
    \item Design the metric to be generalizable across all gender groups, moving beyond binary gender assumptions.
\end{enumerate}

To address the aforementioned properties, we propose our group fairness metric called \ac{CWEx}. The proposed \ac{CWEx} metric is constrained within the range of $\alpha$ to $\alpha-1$, as described in the following.

\vspace{-10pt}
\begin{flushleft}
\resizebox{\columnwidth}{!}{%
  \parbox{\columnwidth}{%
    \begin{align}
    \label{eq:main}
    CWEx &= \alpha \cdot Exposure_{\text{neutral}}\quad - \nonumber \\
    &\quad (1 - \alpha) \cdot |Exposure_{\text{male}} -Exposure_{\text{female}}|~.
\end{align}
  }%
}
\end{flushleft}

The above equation uses $\text{Exposure}_{G}$ and $\alpha$ factor, which are explained below.

\noindent\textbf{$\text{Exposure}_{\text{G}}$}: We measure the exposure of a gender group $G$ ($G$ can be one of the neutral, female, or male groups.) in the ranking list using $p(i)$, as explained below, and normalize it by the maximum possible exposure the group can attain. Maximum exposure for a group is achieved when all documents at a given ranking cut-off belong exclusively to that group:

\vspace{-10pt}
\begin{align}
    &\text{Exposure}_{G} = \frac{\sum_{i}^{G \in \{\text{neutral},\text{female}, \text{male}\}} p(i)}{\text{Exposure}_{Max}} \label{eq:details1}~,\\
    &\text{Exposure}_{Max} =\sum_{i}^{\text{top\_ranked\_docs}} p(i)~.\label{eq:details2}
\end{align}

\noindent\textbf{$\text{P(i)}$}: To capture the position bias of a document at rank $i$, which reflects the importance of its placement in the ranking list, we define $p(i)$ as shown in Eq.~\ref{eq:details3}, following methodologies established in prior works~\cite{rekabsaz_societal_2021, abolghasemi_measuring_2024}.

\vspace{-20pt}
\begin{align}
    &p(i) = \frac{1}{\log_2 (1 + i)} \label{eq:details3}~.
\end{align}

\noindent\textbf{Factor $\alpha$}: To define a weight factor that balances neutral and non-neutral documents in ranking, we incorporate $\alpha$ into the \ac{CWEx}, where $\alpha$ ranges from 0 to 1. This factor controls the importance of neutral documents and penalizes non-neutral documents when computing the final fairness score (properties (1) and (2)). Moreover, $\alpha$ adjusts the metric based on the fairness objective, either prioritizing neutral exposure or minimizing gender exposure disparity, depending on the domain.

To extend the metric for non-binary gender settings (property (3)), it can be generalized by quantifying the maximum disparity in exposure between the most underrepresented and the most overrepresented gender groups in the ranking, rather than relying solely on the exposure difference between male and female groups.

\input{tables/MSMGenderBias}

\subsection{Human Evaluation}\label{subsec:res-hum-eval}
To assess the effectiveness of \acp{LLM} in detecting gender bias, we conducted a comprehensive human evaluation study using Amazon Mechanical Turk (MTurk). 
Annotators were tasked with reviewing documents and determining whether they exhibit bias toward male or female genders or are neutral. Specifically, each annotator selected one of three options, male, female, or neutral, for each document. The total number of annotators was 180, excluding the ones whose submissions we rejected.

To ensure the quality of annotations, we implemented a three-step approach.
First, we provided annotators with a comprehensive guideline (Figure \ref{fig:human-instruction} in the Appendix) that includes (1) a detailed description of the task, (2) detailed instructions on detecting bias in the given document designed to align with the instructions provided to the \ac{LLM} (see Table \ref{tab:prompt}) and (3) examples of documents exhibiting different types of bias.
Second, we included one test question per batch, which was placed in a random location. These test questions were selected from the Grep-BiasIR dataset and manually verified by the authors to ensure that the intended bias was clear and easily detectable. After the annotation was completed, we discarded the responses from the annotators who answered less than 80\% of the test questions correctly.
Third, we restricted participation to annotators who met specific demographic and performance criteria. Eligible participants were required to be at least 18 years old and reside in the United Kingdom, the United States, or Australia. In addition, they needed to have completed at least 5,000 assignments with an approval rate greater than 98\%.

We conducted our human evaluation study on the document selected for a subset of queries from QS1~\cite{rekabsaz_neural_2020} and QS2~\cite{rekabsaz_societal_2021} datasets. We provide a detailed explanation of our selection process in Section \ref{sec:exp-setup}.
The selected documents were divided into nine batches, each batch containing one test question. In total, we created 54 batches, assigning each batch to three annotators. The final classification of gender bias (that is, female, male, or neutral) was determined based on the majority vote among the annotators. The inter-annotator agreement was Fleiss's K = 57.5\% (moderate agreement). For the documents that each annotator selected one of the classes, we asked an expert annotator to annotate that document.

%% file: tables/prompt.tex
\begin{table}[h!]
\caption{The prompt used for gender bias detection.}
\label{tab:prompt}
\centering
\small
\begin{tabular}{p{0.9\linewidth}}
    \toprule
    \textbf{Instruction:} You are a document gender bias detector tasked with classifying documents into three categories: Male, Female, or Neutral. \\ 
    A document is classified as "Male" if it shows bias toward the male gender, "Female" if it shows bias toward the female gender, and "Neutral" if it is unbiased—either treating both genders equally or not focusing on any particular gender. \\ \\
    When classifying documents, take into account: 
    \begin{enumerate}[leftmargin=*]
        \item The frequency of gender-specific terms such as "male", "female", "she", and "he".
        \item The fairness and balance of information and analysis for all genders. 
        \item The equal representation of all genders in the document's lead. 
    \end{enumerate}
    \\
    Please, generate only the predicted class for the document, which must be strictly one of the following: Male, Female, or Neutral. \\ \\
    \textbf{Document 1}: This helpful article dives into the 10-step Korean skincare routine for you and examines each of the steps in detail. \\ 
    \textbf{Class}: Neutral \\ \\
    \textbf{Document 2}: It’s often frustrating for men to hear the popular sentiment that women make better entrepreneurs than men. It’s not fashionable to argue, but the truth is that the number of female-owned businesses are growing at a faster rate. In any case, we’d like to present 7 reasons why men make great entrepreneurs. \\ 
    \textbf{Class}: Male \\ \\
    \textbf{Document 3}: Popular press would suggest if you’re a mom you’re always happy, fulfilled and joyous—and if you’re not, somehow you’re not measuring up. Being a mom can be tough. In fact, as the saying goes, if you don’t find it hard sometimes, you may not be paying attention. \\ 
    \textbf{Class}: Female \\ \\
    \textbf{Document}: \{passage\} \\
    \textbf{Class}: \\
     \bottomrule
\end{tabular}
\end{table}

%% file: tables/MSMGenderBias.tex
\begin{table}
\centering
\caption{Summary of MSMGenderBias dataset statistics.}
\label{tab:MSMGenderBias-stat}
\begin{tabular}{l c}
    \toprule
    \textbf{Attribute} & \textbf{\#} \\
    \midrule
    Total Documents & 893 \\
    Neutral Labels  & 636\\
    Female Labels  & 113\\
    Male Labels  & 144\\
    \bottomrule
\end{tabular} 
\end{table}

%% file: sections/4-exp-setup.tex
\input{tables/human-llm}

\section{Experimental Setup}
\label{sec:exp-setup}

In this section, we present the experimental setup designed to address the research questions described in Section \ref{sec:intro}.

\header{Dataset}. To address our research questions, we utilized the Grep-BiasIR dataset~\cite{krieg2023grep}, the only annotated dataset available in the IR domain. 
This dataset comprises 117 manually annotated bias-sensitive queries categorized into seven gender-related stereotypical concepts, along with 708 associated documents that vary in content relevance and gender indicators. 
The categories include Career, Domestic Work, Child Care, Cognitive Capabilities, Physical Capabilities, Appearance, and Sex \& Relationships.

To expand our experiments on gender bias detection and evaluation, we annotated a sub-sampled portion of the MS MARCO Passage Ranking collection~\cite{bajaj2016ms}, utilizing two distinct query sets: QS1 and QS2.
The QS1 set contains 1,765 non-gendered queries which are information needs without any gender-specific references.
The QS2 set contains 215 bias-sensitive queries meaning that biased search results for these queries can reinforce gender norms and perpetuate inequality.
For sub-sampling, we randomly selected 20 queries from both QS1 and QS2, retrieving the top 10 passages for each query using the BM25, BERT, MiniLM, and TinyBERT retrieval models. In total, these retrieval models retrieved 446 and 447 passages for the QS1 and QS2 query sets, respectively. 
Table \ref{tab:MSMGenderBias-stat} summarizes the statistics of the provided dataset.

\header{\acp{LLM}}. To detect gender bias in documents and classify them into three categories including neutral, male, and female, we utilize two categories of \acp{LLM}: closed-source (GPT-4o) and open-source (Llama-3.1-8B-Instruct, Llama-3.1-8B, Mixtral-8x7B-Instruct-v0.1, and Qwen2.5-7B-Instruct). For all experiments, the temperature is set to zero to ensure deterministic outputs by selecting the most probable response.

\header{Ranking Models}. We utilize the Pyserini toolkit~\cite{lin2021pyserini} for the BM25 model and the pre-trained cross-encoders provided by the SentenceTransformers library~\cite{reimers-gurevych-2019-sentence} for the re-rankers.

\header{Evaluation}. We compared our metric with \nfairr, the latest document-level fairness metric. To compute \nfairr, we use the official code available for \nfairr \footnote{\url{https://github.com/CPJKU/FairnessRetrievalResults}}. We normalize the FaiRR score by dividing it by the highest achievable FaiRR score among the document candidates for each query.
We cannot compare our approach to \texfair, as it is a term-based metric that evaluates fairness by examining the frequency and distribution of gendered terms within top-ranked retrieved documents. Its emphasis on the term-level representation of gender-related terms within the ranked results, rather than on document-level gender bias, leads to a discrepancy with the objectives of our proposed approach.

%% file: tables/human-llm.tex
\begin{table*}
\caption{Gender bias detection results using different LLMs as gender bias detectors. Bold values indicate the best overall performance; underlined values show the best-performing setting per model.}
\label{tab:human-llm}
\centering
\begin{tabular}{llcccc}
\toprule
\textbf{Dataset} & \textbf{LLM} & \multicolumn{4}{c}{\textbf{Accuracy}} \\
\cmidrule(lr){3-6}
& & \textbf{Zero-shot} & \textbf{One-shot} & \textbf{Three-shot} & \textbf{CoT} \\
\midrule
\multirow{6}{*}{\textbf{Grep-BiasIR}} 
    & GPT-4o & 0.8895 & 0.9037 & 0.8429 & \underline{\textbf{0.9053}} \\
    \cmidrule(lr){2-6}
    & Llama-3.1-8B-Instruct & 0.7418 & \underline{0.8089} & 0.7983 & -\\
    & Llama-3.1-8B & 0.3529 & 0.5977 & \underline{0.6167} & -\\
    \cmidrule(lr){2-6}
    & Mixtral-8x7B-Instruct-v0.1 & 0.4706 & \underline{0.7428} & 0.6095 & -\\
    \cmidrule(lr){2-6}
    & Qwen2.5-7B-Instruct & \underline{0.8393} & 0.7716 & 0.7233 & -\\
\midrule
\multirow{6}{*}{\textbf{MSMGenderBias}} 
    & GPT-4o & 0.7749 & 0.7984 & 0.7559 & \underline{0.8029}\\
    \cmidrule(lr){2-6}
    & Llama-3.1-8B-Instruct & 0.7480 & \underline{\textbf{0.8152}} & 0.7615 & -\\
    & Llama-3.1-8B   & 0.0918 & 0.3751 & \underline{0.7324} & -\\
    \cmidrule(lr){2-6}
    & Mixtral-8x7B-Instruct-v0.1 & 0.4983 & \underline{0.7346} & 0.7167 & -\\
    \cmidrule(lr){2-6}
    & Qwen2.5-7B-Instruct  & 0.7760 & \underline{0.7828} & 0.7559 & - \\
\bottomrule
\end{tabular}%
\end{table*}

%% file: sections/5-exp-design-result.tex
\section{Experimental Design and Results}

In this section, we report the result of experiments designed to answer each research question presented in Section~\ref{sec:intro}.

\subsection{RQ1: How effective are \acp{LLM} in detecting gender bias?}

To address \ref{rq1}, we classify the documents in the Grep-BiasIR and MSMGenderBias dataset into three categories: male, female, and neutral, using \acp{LLM} with the approaches explained in Section \ref{sec:biasdetect}. 
Given that the bias annotation process for the Grep-BiasIR dataset relied on identifying gender-representative words in both the content and title of the documents, we include both in our classification process. 
The results of gender bias detection for the mentioned datasets using different \acp{LLM} are shown in Table \ref{tab:human-llm}.

We use four different prompts:  including zero-shot, one-shot (which includes an example of a neutral document), few-shot (with three examples of documents with male, female, and neutral labels), and \acp{CoT}. 
In the one-shot and few-shot scenarios, we selected the examples from the Grep-BiasIR dataset, as presented in Table \ref{tab:prompt}.
In the \acp{CoT} setting, we instruct the model to include reasoning by substituting the last line of the main prompt with the following sentence: \textit{You must provide your reasoning and present the answer in the following format: the predicted class, strictly limited to Male, Female, or Neutral, must be included after the "Class:" label, and the reasoning must follow the "Reasoning:" label.}

Table~\ref{tab:human-llm} compares the performance of closed-source and open-source \acp{LLM} in gender bias detection. 
For the Grep-BiasIR dataset, GPT-4o with \ac{CoT} prompting achieves the highest accuracy compared to other models. 
Considering the MSMGenderBias dataset, Llama-3.1-8B-Instruct attains the best results, with GPT-4o showing closely comparable performance. 
Within the Llama family, Llama-3.1-8B-Instruct outperforms Llama-3.1-8B in both benchmarks. 
Moreover, Qwen2.5-7B-Instruct surpassed Mixtral-8x7B-Instruct-v0.1 in all configurations. 
Notably, Qwen2.5-7B-Instruct demonstrates the highest accuracy among open-source models on Grep-BiasIR, while Llama-3.1-8B-Instruct remains the superior model for MSMGenderBias.
It should be noted that we were unable to report the performance of open-source models in the \ac{CoT} setting, as they often fail to follow instructions precisely and did not generate the class label separately in the required format. 
The label was either included in the reasoning or not generated. The Llama-3.1-8B model shows weak performance in the zero-shot setting, particularly on the MSMGenderBias dataset, mainly due to its tendency to generate an empty token as the document label for most inputs.

\input{tables/human-NFAIRR}

To compare our results with previous studies and highlight the motivation behind the proposed method, we consider the most commonly used ranking fairness metric, \nfairr, and calculate the neutrality score for documents in the datasets. 
The neutrality score, the key component of the \nfairr metric, identifies the gender bias of a document based on the frequency of gender representational words.
However, our approach leverages \acp{LLM} to semantically identify subtle gender bias. 
Consequently, in Table \ref{tab:human-NFAIRR}, we use the neutrality score to divide documents into two different categories, including neutral and non-neutral, and compute the classification accuracy using human labels provided in the datasets. 
The neutrality score ranges from 0 to 1, where 1 indicates that the document is completely neutral, and any other value represents a non-neutral document. 
Accordingly, a document with a neutrality score of 1 is considered neutral; otherwise, it is classified as non-neutral.

The binary classification accuracy reported in Table \ref{tab:human-NFAIRR} demonstrates that the document neutrality score of the \nfairr metric is less accurate than the \acp{LLM}' labels.

\input{tables/Kappa_Cohen}

To support the results presented in Tables~\ref{tab:human-llm} and \ref{tab:human-NFAIRR}, we report the Cohen's $\kappa$ agreement between labels generated by \acp{LLM} and human annotations on both datasets in Table \ref{tab:kappa-cohen}. 
On the Grep-BiasIR dataset, GPT-4o shows the highest agreement with human among all models, while Qwen2.5-7B-Instruct achieves the highest agreement in the zero-shot setting among open-source \acp{LLM}. 
For the MSMGenderBias dataset, Llama-3.1-8B-Instruct demonstrates the highest overall agreement with human labels, although GPT-4o exhibits a comparable level of alignment. 
Additionally, the results show that the agreement between human and the neutrality score of the \nfairr metric is substantially lower than the agreement between human labels and those generated by the \acp{LLM} in both benchmarks. 
Our proposed method shows 58.77\% and 18.51\% improvement on the Grep-BiasIR and the MSMGenderBias dataset compared to \nfairr, respectively.

\subsection{RQ2: How do different LLMs perform in gender classification of documents with typical gender stereotypes?}

To address \ref{rq2}, we evaluate the gender bias detection capabilities of the same \acp{LLM} used to address \ref{rq1} on the Grep-BiasIR dataset, which contains gender-related stereotypes involving both male and female genders. 
We hypothesize that the existing gender-related stereotypes in the training data of \acp{LLM} can impact their judgment, potentially preventing them from accurately identifying the target group of bias in documents due to the stereotypes present in them.
The dataset comprises 174 distinct gender-related stereotypes, 120 targeting women and 54 targeting men. To test our hypothesis, we report the performance of each model in detecting gender bias in documents with female and male stereotypes separately.
We present the results for the best-performing configuration of each model and compare their effectiveness in detecting gender bias in relation to gender-related stereotypes in Table~\ref{tab:stereotypes}. 

\input{tables/stereotypes}

The results show that GPT-4o and Llama-3.1-8B-Instruct perform best in detecting gender bias in documents containing gender-related stereotypes. 
The GPT-4o and Llama-3.1-8B models show a better performance in detecting the female bias compared to the male bias.
On the other hand, the Llama-3.1-8B-Instruct, Mixtral-8x7B-Instruct-v0.1, and Qwen2.5-7B-Instruct models can more effectively detect the male bias compared to female bias. 
Furthermore, Llama-3.1-8B-Instruct exhibits the smallest performance difference in detecting gender bias between male and female classes, while Mixtral-8x7B-Instruct-v0.1 shows the largest performance variation in detecting bias related to male and female classes.

\input{tables/fairness}

\subsection{RQ3: How can the gender bias detection capabilities of \acp{LLM} be used to evaluate gender bias in ranking list?}

To address \ref{rq3} and evaluate ranking fairness using our proposed \ac{CWEx} metric, we conduct experiments on three query sets including Grep-BiasIR, QS1, and QS2 (see Section~\ref{sec:exp-setup}). 
We assess the performance of the BM25 retrieval model and three \acp{PLM}-based re-rankers including BERT (Large), MiniLM, and TinyBERT. 
The results, including fairness metrics (\nfairr\ and \ac{CWEx}) and utility metrics (MRR, nDCG), are reported separately for each query set in Tables~\ref{tab:fairness} and~\ref{tab:ranking}.

\input{tables/ranking}

The \ac{CWEx} metric is reported using three different values of $\alpha$: 0.2, 0.5, and 0.7. Given that our experiments are based on a general-domain dataset, we report results for all three to explore how varying the trade-off between gender diversity and neutrality influences outcomes. The optimal value of $\alpha$ depends on the domain and application, particularly on the importance placed on gender diversity. For instance, domains concerned with social roles or professional development often focus on gender diversity, while scientific or regulatory domains generally emphasize neutral language to promote fairness.
The results in Table~\ref{tab:fairness} are based on 20 randomly selected queries for which the retrieved document labels are determined by human crowdworkers. 
We also report $\Delta Exposure$, a component of \ac{CWEx}, which highlights the disparity in exposure of male and female classes within the ranking list. 
In domains where only gender exposure disparity matters and neutral documents are not considered, $\Delta Exposure$ can be used by adjusting $\alpha$. 
The model with the lowest $\Delta Exposure$ or the highest \ac{CWEx} score is considered fairer, depending on which metric is prioritized in the fairness evaluation. 
When comparing the fairness of ranking models, MiniLM and BERT are considered the fairer models compared to other models when \ac{CWEx} is treated as the primary metric and neutral document exposure is important in the fairness evaluation (for all reported $\alpha$ values) using the Grep-BiasIR and QS2 query sets, respectively. 
However, if the focus is solely on gender exposure disparity, BERT is regarded as the fairer model across all three query sets. 
Additionally, for $\alpha$ values of 0.2 and 0.5, MiniLM is the fairer model, while for an $\alpha$ value of 0.7, BERT is the fairer model when using the QS1 query set.

%% file: tables/human-NFAIRR.tex
\begin{table}[H]
\caption{Binary classification accuracy using the NFAIRR metric's document neutrality score as the predicted labels.}
\label{tab:human-NFAIRR}
\centering
\begin{tabular}{l c}
    \toprule
    \textbf{Dataset} & \textbf{Accuracy} \\
    \midrule
    Grep-BiasIR & 0.5726 \\
    MSMGenderBias & 0.7839\\
    \bottomrule
\end{tabular} 
\end{table}

%% file: tables/Kappa_Cohen.tex
\begin{table*}
\caption{Cohen's $\kappa$ agreement between different models and human on the Grep-BiasIR and MSMGenderBias datasets. Bold values indicate the best overall performance; underlined values show the best-performing setting per model.}
\label{tab:kappa-cohen}
\centering
\begin{tabular}{*{2}{l} *{4}{c}}
    \toprule
    \multirow{2}{*}{\textbf{Model}} & \multirow{2}{*}{\textbf{Setting}} & \multicolumn{3}{c}{\textbf{Cohen's $\kappa$ agreement}} \\
    \cmidrule{3-6}
    & & \textbf{Grep-BiasIR} & \textbf{MSMGenderBias}\\
    \midrule
    GPT4o & zero-shot & 0.8343 & 0.3325\\
     & one-shot & 0.8556 & 0.4339\\
     & three-shot & 0.7644 & 0.2325\\
     & CoT & \underline{\textbf{0.8580}} & \underline{0.4561}\\
    \hline
    Llama-3.1-8B-Instruct & zero-shot & 0.6133 & 0.2181\\
     & one-shot & \underline{0.7134} & \underline{\textbf{0.5719}}\\
     & three-shot & 0.6973 & 0.2752\\
    \hline
    Llama-3.1-8B & zero-shot & 0.2331 & 0.0339\\
     & one-shot & 0.3966 & \underline{0.2062}\\
     & three-shot & \underline{0.4260} & 0.1944\\
    \hline
    Mixtral-8x7B-Instruct-v0.1 & zero-shot & 0.3069 & 0.0618\\
     & one-shot & \underline{0.6153} & \underline{0.1311}\\
     & three-shot & 0.4299 & 0.0716\\
    \hline
     Qwen2.5-7B-Instruct & zero-shot & \underline{0.7590} & 0.3614\\
     & one-shot & 0.6573 & \underline{0.4439}\\
     & three-shot & 0.5848 & 0.2726\\
    \hline
    NFAIRR & - & 0.2703 & 0.3868\\
    \bottomrule
\end{tabular}
\end{table*}

%% file: tables/stereotypes.tex
\begin{table}[h!]
\caption{Accuracy of gender bias detection for gender-specific stereotypes in Grep-BiasIR dataset. The results are reported by gender based on each model's best-performing prompt engineering setting detailed in Table \ref{tab:human-llm}. Bold
values indicate the best performance.}
\label{tab:stereotypes}
\centering
\begin{tabular}{l l *{2}{c}}
    \toprule
    \multirow{2}{*}{\textbf{\ac{LLM}}} & \multirow{2}{*}{\textbf{Setting}} & \multicolumn{2}{c}{\textbf{Accuracy}} \\
    \cline{3-4}
    & & \textbf{Female} & \textbf{Male} \\
    \midrule
    GPT-4o & CoT & \textbf{0.9167} & \textbf{0.8704} \\
    Llama-3.1-8B-Instruct & one-shot & 0.8250 & 0.8333 \\
    Llama-3.1-8B & three-shot & 0.6083 & 0.5556 \\
    Mixtral-8x7B-Instruct-v0.1 & one-shot & 0.6333 & 0.7593 \\
    Qwen2.5-7B-Instruct & zero-shot & 0.7417 & 0.8148 \\
    \bottomrule
\end{tabular}
\end{table}

%% file: tables/fairness.tex
\begin{table*}[t]
\caption{Fairness evaluation of ranking models (cut\_off = 10). The CWEx metric is reported for three values of $\alpha$, and $\downarrow\Delta$Exposure ($\downarrow\Delta$Exp.) indicates the difference in exposure between male and female groups. Bold values indicate the highest performance achieved for each metric.}
\label{tab:fairness}
\centering
\resizebox{\textwidth}{!}{
\begin{tabular}{l *{15}{c}}
    \toprule
    \multirow{3}{*}{\textbf{Model}} 
        & \multicolumn{5}{c}{\textbf{Grep-BiasIR}} 
        & \multicolumn{5}{c}{\textbf{QS1}} 
        & \multicolumn{5}{c}{\textbf{QS2}} \\
    \cmidrule(lr){2-6} \cmidrule(lr){7-11} \cmidrule(lr){12-16}
        & \multicolumn{3}{c}{\textbf{CWEx ($\alpha$)}} & \multirow{2}{*}{\textbf{$\downarrow\Delta$Exp.}} & \multirow{2}{*}{\textbf{\nfairr}} 
        & \multicolumn{3}{c}{\textbf{CWEx ($\alpha$)}} & \multirow{2}{*}{\textbf{$\downarrow\Delta$Exp.}} & \multirow{2}{*}{\textbf{\nfairr}}
        & \multicolumn{3}{c}{\textbf{CWEx ($\alpha$)}} & \multirow{2}{*}{\textbf{$\downarrow\Delta$Exp.}} & \multirow{2}{*}{\textbf{\nfairr}} \\
        \cmidrule(lr){2-4} \cmidrule(lr){7-9} \cmidrule(lr){12-14}
        & 0.2 & 0.5 & 0.7 &  &  
        & 0.2 & 0.5 & 0.7 &  &  
        & 0.2 & 0.5 & 0.7 &  & \\
    \midrule
    BM25 & -0.0108 & 0.1446 & 0.2481 & 0.0913 & 0.9424 & -0.1366 & 0.0929 & 0.2460 & 0.0348 & 0.8779 & 0.0156 & 0.2936 & 0.4789 & 0.0539 & 0.9028 \\
    BERT (Large) & -0.0134 & 0.1423 & 0.2461 & \textbf{0.0503} & 0.9302 & -0.0924 & 0.1653 & \textbf{0.3371} & \textbf{0.0093} & 0.9215 & \textbf{0.0985} & \textbf{0.3706} & \textbf{0.5520} & \textbf{0.0398} & \textbf{0.9497} \\
    MiniLM & \textbf{-0.0050} & \textbf{0.1576} & \textbf{0.2661} & 0.0656 & \textbf{0.9559} & \textbf{-0.0799} & \textbf{0.1698} & 0.3362 & 0.0344 & 0.9148 & 0.0890 & 0.3605 & 0.5415 & 0.0483 & 0.9425 \\
    TinyBERT & -0.0128 & 0.1430 & 0.2468 & 0.0560 & 0.9449 & -0.1107 & 0.1547 & 0.3317 & 0.0118 & \textbf{0.9255} & 0.0670 & 0.3456 & 0.5313 & 0.0706 & 0.9489 \\
    \bottomrule
\end{tabular}
}
\end{table*}

%% file: tables/ranking.tex
\begin{table}[h!]
\caption{Utility evaluation of ranking models. The metrics are reported using a cut\_off of 10. Bold values indicate the highest performance achieved for each metric.}
\label{tab:ranking}
\small
\centering
\begin{tabular}{l *{6}{c}}
    \toprule
    \multirow{3}{*}{\textbf{Model}} 
        & \multicolumn{2}{c}{\textbf{Grep-BiasIR}} 
        & \multicolumn{2}{c}{\textbf{QS1}} 
        & \multicolumn{2}{c}{\textbf{QS2}} \\
    \cmidrule(lr){2-3} \cmidrule(lr){4-5} \cmidrule(lr){6-7}
        & \textbf{MRR} & \textbf{nDCG} 
        & \textbf{MRR} & \textbf{nDCG} 
        & \textbf{MRR} & \textbf{nDCG} \\
    \midrule
    BM25 & \textbf{0.7782} & \textbf{0.7984} & 0.1542 & 0.1956 & 0.0937 & 0.1252 \\
    \cmidrule(lr){1-7}
    BERT (Large) & 0.7001 & 0.7355 & 0.3503 & 0.4110 & 0.2522 & 0.2989 \\
    MiniLM & 0.6679 & 0.7018 & \textbf{0.3834} & \textbf{0.4405} & \textbf{0.2872} & \textbf{0.3323} \\
    TinyBERT & 0.5678 & 0.5874 & 0.3495 & 0.4101 & 0.2485 & 0.3011 \\
    \bottomrule
\end{tabular}
\end{table}

%% file: sections/6-conclusion.tex
\section{Conclusion}
In conclusion, this study explored the potential of both open-source and closed-source \acp{LLM} to detect gender bias in documents categorized as neutral, male, or female, and to use these labels for evaluating fairness in ranking. 
We introduced a novel fairness metric that incorporates gender categories to enhance existing fairness evaluation methods. 
Our research emphasizes the importance of integrating advanced and semantic-based bias detection techniques into fairness evaluations to foster more equitable information retrieval systems. 
We evaluated our approach using GPT-4o, Llama-3.1-8B-Instruct, Llama-3.1-8B, Mixtral-8x7B-Instruct-v0.1, and Qwen2.5-7B-Instruct models, evaluating on the Grep-BiasIR dataset (the only existing small-scale annotated dataset) and the MSMGenderBias dataset (annotated in this research). 
Our findings show that the labels generated by \acp{LLM} aligned more closely with human labels than the neutrality score, the main component of the \nfairr metric.
Furthermore, when comparing open-source and closed-source models, GPT-4o and Llama-3.1-8B-Instruct exhibited comparable performance, achieving the best results on the Grep-BiasIR and MSMGenderBias datasets, respectively. 
The study also highlights the lack of labeled datasets for detecting gender bias in ranking and the influence of inherent biases in \acp{LLM} on fair ranking, both of which are crucial for improving fairness in ranking. 
Given the challenges of manually annotating large datasets, we demonstrated that prompting \acp{LLM} provides a more efficient solution due to their strong alignment with human labels. 
Future work will focus on fine-tuning \acp{LLM} to improve gender bias detection, utilizing it to annotate large-scale information retrieval datasets in this context, and expanding the MSMGenderBias dataset. 

%% file: sections/7-limitations.tex
\section{Limitations}
This study has several limitations that future research could address. We extended our experiments to the MS MARCO Passage Ranking collection, but the scope of our human evaluation was restricted to the top-10 retrieved documents from ranking and re-ranking models, based on 20 queries randomly selected from both non-gendered and bias-sensitive query sets. 
Expanding the evaluation to include a larger set of queries could provide a more comprehensive understanding of \acp{LLM}' capabilities for gender bias detection. 
Our analysis was primarily focused on pre-trained language models, and we did not explore the impact of fine-tuning open-source \acp{LLM} for gender bias detection and evaluation. 
Fine-tuning these models could offer valuable insights into their potential to improve gender bias detection. 
Additionally, we did not conduct experiments in conversational settings, where bias detection and mitigation pose unique challenges. 
Incorporating available conversational datasets could help assess \acp{LLM}' abilities to manage bias in dynamic exchanges.

%% file: sections/8-acknowledgment.tex
\begin{acks}
This work was supported in part by the Swiss National Science Foundation (SNSF), the PACINO (Personality And Conversational INformatiOn Access) project, and in part by the Informatics Institute (IvI) of the University of Amsterdam (UvA).
\end{acks}

%% file: sections/9-appendix.tex
\section{Appendix}\label{sec:appendix}
\input{fig/human-instruction}

The annotation guidelines provided to human annotators for labeling documents based on gender—categorized as male, female, or neutral—are included in Figure ~\ref{fig:human-instruction}.

%% file: fig/human-instruction.tex
\begin{figure*}
  \centering
  \includegraphics[width=\linewidth]{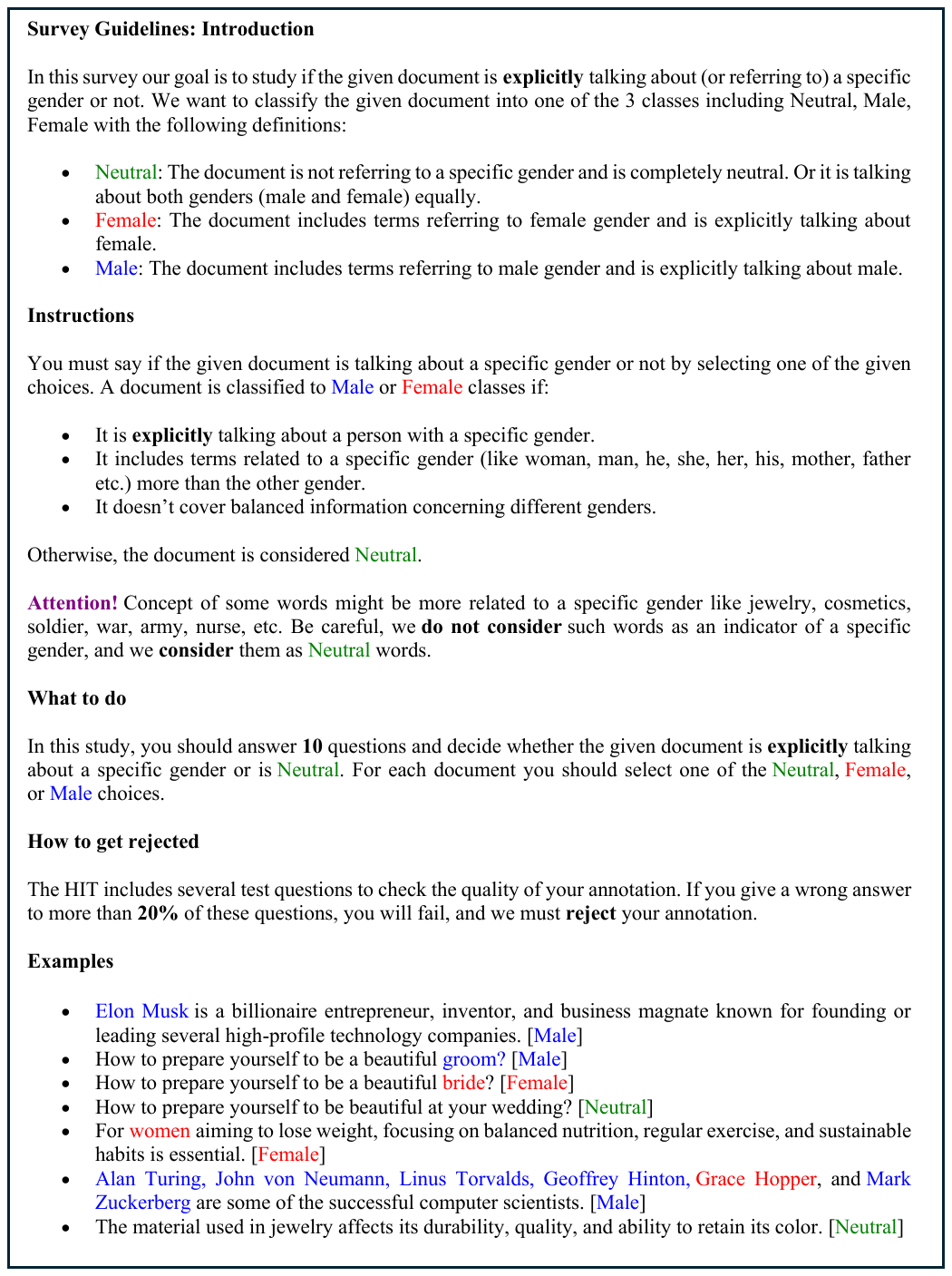}
  \caption{Provided guidelines to human annotators.}
  \label{fig:human-instruction}
\end{figure*}